\documentclass[12pt]{iopart}
\begin{document}

\vspace*{-7mm}

\hfill MZ-TH/98-62

\hfill OUNP-99-03

\title[Standard Model Large-$E_T$ Processes and Searches at HERA]%
{Standard Model Large-$E_T$ Processes \\
and Searches for New Physics at HERA}

\author{M W Krasny and H Spiesberger}
    
\address{\dag Balliol College and Nuclear and Astrophysics
      Laboratory, Keble Road, Oxford OX13RH, UK and IN2P3-CNRS,
      Universit\'es Paris VI et VII, 4 pl Jussieu, T33-RdC, 75252 Paris,
      France}
    
\address{\ddag Institut f\"ur Physik,
      Johannes-Gutenberg-Universit\"at, D-55099 Mainz, Germany}

\begin{abstract}
  Existing and missing calculations of standard model processes
  producing large transverse energy in electron-proton interactions at
  HERA are reviewed. The adequacy of the existing standard model Monte
  Carlo programs for generic searches of exotic processes is
  analyzed\footnote[3]{Contribution to the 3rd UK Phenomenology Workshop
    on HERA Physics, September 1998, Durham}.
\end{abstract}


\section{Introduction}

The two experiments H1 and ZEUS at HERA have accumulated over the last
six years almost 100 pb$^{-1}$ of luminosity, providing the first
glimpse at the short distance frontier of lepton-nucleon interactions.
A substantial effort in analyzing the data has been focused on extending
the rejection limits for parameters of various known extensions of the
standard model (SM) \cite{exp-searches}.  In the corresponding {\it
  dedicated searches} both the data selection and the analysis method
were optimized to detect the {\it anticipated} experimental signatures
of these extensions.
 
Another approach, which we would like to call {\it generic searches},
consists of analyzing a broad class of events characterized by the
presence of a large transverse energy particle or a system of particles
used as tags for short distance and/or large mass processes
\cite{H1generic}.  The event selection criteria for such searches are
optimized to cover those phase space regions where the standard model
predictions are sufficiently precise to detect anomalies rather than to
cover the phase space region where anomalies are expected.  They allow
one to detect anomalies in the {\it relative} abundance of several
processes.  In general, generic searches minimize the chances that
unexpected novel phenomena are overlooked if their manifestations in the
data do not follow one of the predefined scenarios and enlarge the
discovery potential if signatures of new phenomena are weak but present
in more than one final state topology.

Both dedicated and generic searches rely on the knowledge of standard
model predictions for the distributions of events over the available
phase space.  This requires dedicated SM calculations and their
implementations in Monte Carlo event generators. In our view, these
indispensable theoretical tools for searches of novel phenomena at HERA
need upgrading.  The lack of tools which are applicable over the full
phase space leaves an important fraction of registered data un-analyzed.
The high-energy, i.e.\ high-$y$, frontier of $\gamma p$ scattering is an
example of a phase space region unique to HERA which was largely
untouched by dedicated searches. In this region both deep inelastic and
photoproduction processes contribute and a tedious procedure of matching
the respective calculations and their Monte Carlo implementations is
required \cite{H1generic}. In dedicated searches, leaving out a fraction
of the phase space has only weak impact on the derived exclusion limits.
For generic searches, the development of appropriate theoretical tools
covering the full phase space, in particular close to its boundaries, is
mandatory if one wants to maximize the efficiency for detecting
unexpected novel phenomena rather than to establish the rejection limits
for the expected ones.

Standard model predictions for many important processes and
corresponding Monte Carlo event generators to simulate the bulk of
events in $ep$ scattering are available. These programs may, however,
suffer from approximations which could turn out to be severe in corners
of the phase space or for specific final states.  As an example we
mention the fact that the {\tt DJANGO} branch of the Monte Carlo program
{\tt DJANGOH} \cite{DJANGOH} is well-suited for the simulation of the
inclusive DIS cross section, but does not take into account effects due
to the emission of photons from quarks. For the total cross section this
is a reasonable approximation, but is not acceptable when searching for
final states containing photons in the vicinity of jets.  An example
where this is relevant is the search for excited quarks.

The precision of tools for searches is of a different quality compared
to that for a precision measurement of structure functions or parameters
of the standard model.  In general the statistical significance of novel
phenomena is weak and a precision of the order of 20\,\% matches already
the statistical and systematic errors of the measurement. Therefore,
electroweak radiative corrections are not crucial and pure QED
corrections are sufficiently accurate when calculated in the leading
logarithmic approximation.  QCD corrections, however, need to be
controlled.

In some cases significant contributions may arise from processes which
are not accessible to perturbative methods, or which are sensitive to
soft-gluon resummation.  An example for the first case is $ep
\rightarrow \gamma X$ which receives contributions from $ep \rightarrow
ep\gamma$ in the absence of cuts on the hadronic final state
\cite{comptons}, an example for the latter case is the production of two
jets with equal $p_T$ \cite{twojetg}.  Therefore, the scope of the
calculations needed for generic searches makes them much more ambitious
than in cases where cuts are applied to remove the ``difficult'' phase
space regions.
 
The main requirement for the theoretical tools used in searches is their
completeness: all standard model processes contributing to a particular
final state have to be controlled.  In determining a complete list of
all relevant processes, a close collaboration between theorists and
experimentalists is needed because measurement effects significantly
enlarge the number of processes which could potentially contribute to a
particular final state topology. For example, the production of jets,
which could occasionally mimic electrons, muons, taus, photons and, if
undetected, neutrinos may contribute to the majority of the final states
considered below.  The theoretical tools in this case should not only
provide the means to control the jet production rate and their spectra
but, in addition, to control the probabilities for a jet to fragment
into rare particle topologies, like single tracks, jets with only
neutral particles, etc. The discussion of these latter aspects is beyond
the scope of this paper.

In this note we provide a short review of the standard model
calculations and their Monte Carlo implementations which can be used in
searches for new physics and point out areas where progress is needed.
The discussion is organized according to a possible classification of
generic searches: inclusive single-particle spectra for electrons,
neutrinos, jets, photons and so on, and spectra for various combinations
of these particles are considered in turn.


\section{Inclusive single-particle spectra}
 
\subsection{Electrons and Neutrinos}

Measurements that only look for electrons or neutrinos (i.e., missing
transverse momentum) in the final state correspond to the classical
inclusive deep inelastic scattering. The precision of standard model
predictions is determined by that of the structure functions input.
These are obtained from global fits to a wide variety of data taking
into account $Q^2$ evolution according to perturbative QCD in
next-to-leading order. Theoretically, the predictions are on a sound
basis since the operator product expansion approach for the inclusive
measurement is well-founded. Therefore the precision is limited by the
experimental accuracy of the data that are used in the fits. Recent
estimates suggest precisions well below 10 per cent \cite{WG1} including
uncertainties from scale variation and $\alpha_s$. For the longitudinal
structure function $F_L$, where presently available measurements are not
very precise, QCD predictions can be used. The inclusive measurement is
sensitive to the simulation of hadronic final states only to the extent
that the latter modify the reconstructed kinematical variables and enter
the determination of experimental corrections.

Electroweak radiative corrections are known to order $\alpha$ and can be
taken into account in the Monte Carlo program {\tt HERACLES}
\cite{heracles}.  Only small residual uncertainties are expected at
large $Q^2$.  One should, however, keep in mind that taking into account
radiative corrections requires the knowledge of structure functions down
to $Q^2 = 0$.  Therefore, additional uncertainties from low-$Q^2$
structure functions are present, although only at the level of a few per
cent.

\subsection{Jets}

Jets in high-energy scattering are a genuine QCD testing ground and have
therefore received much attention in theoretical calculations: NLO
calculations were completed for both $(1+1)$- and $(2+1)$-jet production
in deep inelastic scattering \cite{jets1} and their implementations in
Monte Carlo programs \cite{DISJET,PROJET} have been available already
for some time. Jet cross sections need the definition of a jet finding
algorithm and the older calculations were restricted to the modified
JADE scheme. More recent progress in Refs.\ 
\cite{MEPJET,DISENT,DISASTER,JETVIP} allows jet cross sections to be
calculated with more general jet definitions\footnote{A comparison of
  the programs {\tt MEPJET} \cite{MEPJET}, {\tt DISENT} \cite{DISENT}
  and {\tt DISASTER++} \cite{DISASTER} revealed discrepancies, see
  \cite{DISASTER} for details.}. As shown in \cite{PoSe} the choice of
the jet algorithm has considerable influence on the size of scale
uncertainties.  $(3+1)$-jet cross sections are known to leading order
only and implemented in the available programs
\cite{MEPJET,DISENT,DISASTER,JETVIP}, the LO $(4+1)$-jet cross section
also in \cite{MEPJET}.

Jets arising from photoproduction have also been investigated and NLO
calculations have been available for many years \cite{jetgp}. Recent
work \cite{jetgp2,kkk} has improved the understanding of various aspects
like scale uncertainties and the matching of theoretical and
experimental jet definitions. The efforts of Klasen, Kramer and P\"otter
\cite{jetlowq} have lead to NLO predictions for $(2+1)$-jet production
also in the transition region from photoproduction to deep inelastic
scattering. The results were implemented recently in the Monte Carlo
generator {\tt JETVIP} by P\"otter \cite{JETVIP}.

The theoretical tools for generic searches in jet production need
further improvements. The most important restriction of presently
available programs is related to very large $Q^2$ where $Z$ exchange
contributions (and $W$ exchange in charged current scattering) are
important: these contributions are available only at leading order in
the general-purpose programs like {\tt LEPTO} \cite{LEPTO} or {\tt
  DJANGOH} \cite{DJANGOH}.  Other possible improvements are related to
the treatment of heavy quarks, resolved contributions and, for future
experiments, polarized beams. We refer to the accompanying reports
\cite{WG3,hadig} of these proceedings for more details.

\subsection{Photons}

Searching for a photon in the final state without requiring additionally
a large $p_T$ electron or neutrino, one has to cope with contributions
from photoproduction (or more generally very low $Q^2$), deep inelastic
scattering (i.e., large $Q^2$) and the transition region connecting
these two cases. For the two extreme situations, NLO calculations are
available, for photoproduction in Ref.\ 
\cite{isogammaph1,isogammaph2,isogammaph3} and for DIS in Ref.\ 
\cite{isogammadis}\footnote{However, not including $Z$ exchange.}. The
transition region, however, has not yet been investigated.

With the present situation, the calculations for DIS and photoproduction
have to be combined by hand with all the difficulties arising when
approaches of different authors using different conventions are to be
matched. In principle, a cut on $Q^2$ should allow one to separate the
two cases: the calculation in the DIS region is used for $Q^2$ above a
lower limit $Q^2_{\rm min}$ of the order of a few GeV${}^2$ and the
results for photoproduction can be folded with the Weizs\"acker-Williams
spectrum of photons originating from the incoming lepton. $Q^2_{\rm
  min}$ enters the normalization of the Weizs\"acker-Williams spectrum.
In practice this matching was never done and progress similar to that 
which resulted in the Monte Carlo program {\tt JETVIP} would be very
useful to avoid this difficulty.

Similarly to the case of radiative corrections to inclusive DIS, there
is a contribution from virtual $e\gamma^{\ast}$ Compton scattering
producing a photon with large transverse momentum (balanced by the
electron $p_T$) but no large momentum scale in the hadronic subprocess
$\gamma^{\ast} p \rightarrow X$ \cite{comptons,courauk}. Substantial
contributions both from quasi-elastic scattering, i.e.\ with $ep
\rightarrow ep\gamma$, as well as inelastic scattering with low-mass
hadronic final states (deep-inelastic Compton process) $ep \rightarrow
e\gamma X$ are expected. The first case can be described with the help
of the well-known formfactors for elastic $ep$ scattering and
corresponding predictions can be obtained with the help of the programs
{\tt HERACLES} or {\tt COMPTON} \cite{courau}; the latter case is less
well understood and needs further improvements.

\subsection{$W$ and $Z$}

The Monte Carlo generator {\tt EPVEC} based on the calculations by Baur,
Vermaseren and Zeppenfeld \cite{Baur}\footnote{For previous calculations
  see the references in \cite{Baur}.} has been used in searches for
anomalous $W$ couplings \cite{David_ZEUS} and as a tool to control the
$W$ and $Z$ contribution to isolated lepton production \cite{H1lepton}.
{\tt EPVEC} is presently being checked against an independent
calculation of Dubinin and Song \cite{Dubinin}.  As in the case of
inclusive photon and inclusive jet production the main difficulty in the
calculation of NLO corrections boils down to matching the DIS and the
photoproduction contributions. In Ref.\ \cite{Baur} this matching is
defined in terms of the virtuality of a quark exchanged in the
$u$-channel which introduces a cutoff $u_{\rm min}$ and uncertainties
related to variations of this unphysical parameter. A better matching
scheme would separate the deep inelastic regime from photoproduction in
terms of the virtuality of the exchanged photon.  Remaining $u$-pole
singularities can be absorbed into the parton distribution functions in
the photon, giving rise to large QCD corrections for the resolved
contribution. First numerical results of a corresponding calculation
by Nason, R\"uckl and Spira were reported on this workshop \cite{Spira}.
When finalized, a Monte Carlo implementation of this calculation will
allow to obtain the $W$ total cross sections, as well as the spectra at
high transverse momentum, with sufficient precision to be useful in
searches for anomalies. The spectrum at low $p_T$ of the $W$ might
require resummation of soft gluon contributions.

\subsection{Muons and taus}

Large transverse momentum muons and taus are produced at HERA
predominantly as decay products of $W$ and $Z$ bosons. Their production
has been discussed in the previous subsection.  A comparably large
contribution is due to non-resonant dimuon (ditau) production.
Leading-order diagrams in photon-photon interactions constituting a
subset of these processes have been calculated in Ref.\ \cite{Vdilepton}
and subsequently implemented in the Monte Carlo program {\tt LPAIR}
\cite{LPAIR}. {\tt LPAIR} is restricted to the case of large invariant
masses of the lepton pair.  It is not complete for searches based on a
single lepton tag because processes of internal conversion of virtual
photons emitted by the quark or by the electron are not included. These
processes are dominant for dimuons (ditaus) produced with low invariant
mass, and comparable in size to photon-photon interactions at high
transverse momentum. The missing contributions have been calculated in
Ref.\ \cite{Bdilepton} and implemented subsequently in the Monte Carlo
program {\tt TRIDENT} \cite{TRIDENT}.  This program was, however, not
used so far in experimental analyses and additional testing seems to be
required. A new Monte Carlo program, called {\tt GRAPE-Dilepton}, is
presently being developed on the basis of the {\tt GRACE} system
\cite{abe}. It will take into account elastic and quasi-elastic
contributions and the complete matrix elements needed in the deep
inelastic regime as well as simulation of the hadronic final state.  No
NLO calculations are available at present to control theoretically the
inclusive spectra of muons and taus at HERA at the level of precision
required for searches.  Progress in this domain has to include a Monte
Carlo program implementing all dilepton production processes, including
diagrams with on- and off-shell $Z^0$'s.


\section{Multi-body final states}

Multi-body final states, especially those involving an unconventional
particle composition, are of particular interest for HERA generic
searches. When trying to reveal novel phenomena in multi-body final
states, the advantage due to higher energies at the Tevatron is balanced
by significantly smaller QCD backgrounds at HERA.

NLO calculations are much more involved for multi-body final states and
available only for a few cases. Even LO calculations are not worked out
for all interesting processes. In such cases the general-purpose program
packages {\tt CompHEP} \cite{comphep} or {\tt GRACE} \cite{grace} may be
helpful to obtain good estimates. It has to be stressed however that
additional work will be needed, for example to take into account the
effects due to the hadronization of final states.

Searches for anomalies in the production of electron pairs and triplets
or photon pairs, as well as those looking for $e \mu$, $e \tau$, $\mu
\mu$, $\tau \tau$ $e \mu \mu$ and $e \tau \tau$ cannot rely at present
on complete standard model predictions. We refer here to the discussion
in section 2.5.

Searches for the anomalous production of an $e \gamma$ system made so
far were limited to that fraction of the available phase space which can
be controlled by the Monte Carlo programs {\tt DJANGOH}\footnote{{\tt
    DJANGOH} provides an option which allows simulation of the hadronic
  final state but does not include quarkonic radiation ({\tt DJANGO}
  branch) or an option including quarkonic radiation but without
  hadronization ({\tt HERACLES} branch).} and {\tt COMPTON}. More recent
calculations \cite{isogammadis} include QCD corrections to order
$O(\alpha_s)$.  Although these calculations can not easily be combined
with more complete Monte Carlo generators, they are useful for
comparisons with data corrected for detector effects and give a
theoretical handle for studies of the $e\gamma$ and $e\gamma+$jet
systems.  Similar calculations for the production of $\nu \gamma$ and
$\nu \gamma+$jet final states exist in LO but remain to be done in NLO.

Significant progress has been made recently in the precision of the
standard model calculations for $e+$jet and $e+2$jets final states. We
refer here to the discussion presented in section 2.2.\ and to the
contribution of P\"otter and Seymour in these proceedings. As already
mentioned, the corresponding calculations for final states with $\nu
+$jet and $\nu+2$ jets remain to be done.

The searches for anomalies in the production of final states containing
$e \nu$, $\mu \nu$, $\tau \nu$ as well as $e \nu+$jet, $\mu \nu+$jet,
$\tau \nu+$jet, rely on calculations for $W$ production which was
discussed in section 2.4.  A particular need for calculations of QCD
corrections for three body final states involving large $E_T$ jets has
to be underlined in the context of the reported observation of anomalous
events of this type in the H1 data \cite{H1lepton}.  Other multi-body
final states where QCD corrections have not yet been calculated are
$\mu+$jet, $\tau+$jet, 3jets, and many more rare final states.

In several multi-particle final states where theoretical control is
poor, generic searches can, at present, only be based on the relative
abundance of various processes. Simple approximate formulas which, for
example, relate the cross section for lepton pair or jet pair production
to those for large-$E_T$ photon production are known (see for example
\cite{srelation}) and expected to be precise enough for the present
requirements.


\section{Conclusions}

The forseen upgrade of the HERA machine is expected to lead to an
increase of the collected luminosity by at least a factor of ten by the
year 2005. This improvement deserves being matched by corresponding
upgrades of theoretical tools for generic searches in order to fully
exploit the search potential at HERA.  If deviations from the standard
model predictions are observed, a combined treatment of different
multi-particle final states is expected to help in understanding the
anomalies. For example anomalies in the internal structure of the proton
at short distances may reveal themselves by looking at the spectra of
jets associated with various multi-body final states \cite{H1generic}.
Generic searches are helpful in interlinking the experimental aspects of
physically different processes.  This facilitates the assessment of the
size of systematic measurement errors.


\ack We acknowledge discussions on various aspects of standard model
calculations with T.\ Abe, R.\ Devenish, D.\ Graudenz and B.\ P\"otter.


\end{document}